\documentclass[10pt,twocolumn,letterpaper]{article}

\usepackage{cvpr}
\usepackage{times}
\usepackage{epsfig}
\usepackage{graphicx}
\usepackage{amsmath}
\usepackage{amssymb}
\usepackage{multirow}

\usepackage[pagebackref=true,breaklinks=true,letterpaper=true,colorlinks,bookmarks=false]{}

 \cvprfinalcopy 

\ifcvprfinal\fi
\begin{document}

\title{Automatic detection of passable roads after floods in remote sensed and social media data}

\author{Kashif Ahmad\\
Adapt Centre, School of Computer Science and Statistics \\ Trinity College Dublin,  Ireland \\
{\tt\small ahmadka@tcd.ie}
\and
Konstantin Pogorelov\\
Simula Research Laboratory \\University of Oslo, Norway\\
{\tt\small konstantin@simula.no}
\and
Michael Riegler\\
Simula Metropolitan Center for Digital Engineering\\ University of Oslo, Norway\\
{\tt\small michael@simula.no}
\and
Olga Ostroukhova\\
Research Institute of Multiprocessor Computation Systems \\ n.a. A.V. Kalyaev, Russia\\
{\tt\small konstantin@simula.no}
\and
P{\aa}l Halvorsen\\
Simula Research Laboratory \\University of Oslo, Norway\\
{\tt\small paalh@ifi.uio.no}
\and
Nicola Conci\\
University of Trento, Italy\\
{\tt\small nicola.conci@unitn.it}
\and
Rozenn Dahyot \\
Adapt Centre, School of Computer Science and Statistics \\ Trinity College Dublin,  Ireland \\
{\tt\small DAHYOTR@tcd.ie}
}

\maketitle

\begin{abstract}
   This paper addresses the problem of floods classification and floods aftermath detection utilizing both social media and satellite imagery. Automatic detection of disasters such as floods is still a very challenging task. The focus lies on identifying passable routes or roads during floods. Two novel solutions are presented, which were developed for two corresponding tasks at the MediaEval 2018 benchmarking challenge. The tasks are (i) identification of images providing evidence for road passability and (ii) differentiation and detection of passable and non-passable roads in images from two complementary sources of information. For the first challenge, we mainly rely on object and scene-level features extracted through multiple deep models pre-trained on the ImageNet and Places datasets. The object and scene-level features are then combined using early, late and double fusion techniques. To identify whether or not it is possible for a vehicle to pass a road in satellite images, we rely on Convolutional Neural Networks and a transfer learning-based classification approach. The evaluation of the proposed methods are carried out on the large-scale datasets provided for the benchmark competition. The results demonstrate significant improvement in the performance over the recent state-of-art approaches.
\end{abstract}

\section{Introduction}

Natural disasters, such as floods, earthquakes and storms, may cause significant damage to both human life and infrastructures. In such adverse events, an instant access to relevant information may certainly help in the rescue operations, which will ultimately help in mitigating the damage \cite{ahmad2018social,ahmad2017jord}. 
Having an idea of the scope of damage inflicted due to a disaster, government and non-government organizations could allocate their resources to the affected areas accordingly. In such situations, especially in flood events, information about the passability of the roads, i.e., it is possible to travel through the affected areas, is a crucial element for emergency response and deployment of the resources for rescue operations.

In this respect, social media emerged as an important source of information and has been proved very effective in emergency situations, where news agencies could not provide information at all or in time \cite{brouwer2017probabilistic,yin2012using}. For instance, in \cite{stelter2008citizen}, several situations have been reported where news agencies with conventional sources of information failed to report in a timely fashion simply due to the insufficient number of reporters covering the world. On the other hand, due to wide geographical coverage and high spatial and multi-spectral resolutions, satellite imagery has been widely used for the analysis of natural disasters and their impact on the environment \cite{kansas2016using,lagerstrom2016image}. 

The joint use of social media and remotely sensed information has been already investigated in the literature for the analysis of natural disasters and their potential impact on the environment \cite{ahmad2018social,bischke2017detection,jing2016flood,nogueira2018exploiting}.
To encourage research in this area, as well as to compare the performance of different software solutions, a task at the MediaEval benchmarking initiative has been initiated with particular attention to flood detection for two consecutive years i.e., MediaEval~2017 \cite{bischke2017multimedia} and MediaEval~2018 \cite{bischke2018multimediasatellite}. 
In MediaEval 2017, the challenge aimed at flood detection in images from social media and satellite. In Mediaeval 2018, the challenge targeted analysis of social media and satellite imagery for detection of passable roads in flood-affected regions.
The 2018 challenge is composed of two parts: 
\begin{enumerate}
    \item \label{item:FCSM}\textbf{FCSM: Flood Classification for Social Multimedia}. This task is further divided in two sub-tasks aiming to predict: 
            \begin{enumerate}
        \item  whether there are evidences of a flood in a given social media image or not and,
        \item  if evidences of flood exists in the image, whether it is possible to pass through the flooded road (passability).
    \end{enumerate} 
    \item \label{item:FDSI} \textbf{FDSI: Flood Detection in Satellite Imagery}.  This task aims to analyze the roads from satellite images, and predict whether or not it is possible for a vehicle to pass a road.
    \end{enumerate}
 This paper addresses the MediaEval 2018 challenge. 
For FCSM (\ref{item:FCSM}), we rely on an ensemble of several deep models using three different fusion techniques. For FDSI (\ref{item:FDSI}), we rely on a Convolutional Neural Network (CNN) architecture and a transfer learning-based classification approach to identify passable roads in satellite imagery.
The main contributions of the work can be summarized as:
\begin{itemize}
 \item[(i)]  We analyze flood-related images from social media and satellite to identify passable roads. In details, we aim to analyze (a) whether the images provide evidence for road passability and (b) to differentiate between images showing passable vs non passable roads. 
 \item[(ii)] On thhe FCSM task, being an important ingredient of several multimedia analysis applications frameworks, we analyze the performances of four models from three well-known architectures, pre-trained on object and places datasets, both individually and in different combinations using three different fusion techniques, namely early, late and double fusion. We believe such rigorous analysis will set a benchmark for the future research on the topic.  
  \item[(iii)] For the analysis of satellite imagery in the FDSI task, we propose a CNN- and a transfer learning-based classification approach to identify passable roads in satellite imagery. 
  \item[(iv)] We have performed experiments on a challenging benchmark dataset provided for the benchmark competition, and show that better scores are achieved compared to the recent literature.
\end{itemize}
The rest of the paper is organized as follows. Related work is discussed in Section \ref{s:relwork}, followed by a detailed presentation of our proposed solution (Section \ref{sec:proposed}). We show experimentally how our approach performs (Section \ref{sec:results}), and we conclude and provide potential future research directions in Section \ref{sec:conclusion}.

\section{Related Work}
\label{s:relwork}

The literature on natural disasters detection and analysis can be roughly divided into two parts, namely (i) disaster detection in social media, and (ii) disaster detection in satellite imagery. In the next subsections, we provide a detailed review of the relevant literature in both domains.

\subsection{Disasters detection in Social Media Images}

In recent years, several applications have emerged to make use of data posted on social media platforms in combination to other media streams available (e.g. Google Street view imagery, OpenStreetMap map) allowing 3D reconstruction of cities \cite{Seitz2008} and automatic discovery and geo-tagging of objects \cite{krylov2018automatic}. 

Geo-located and time-stamped data available in form of text and visual content on social media have also been widely utilized for disaster events analysis to gather useful information to be used in rescue and rehabilitation \cite{ahmad2018social}.
To this aim, most of the approaches rely on two types of complementary information including visual contents and the additional information associated with images in the form of metadata, such as user tags, geo-location and temporal information. For instance, in \cite{yang2011hierarchical}, users' tags and other useful information from metadata are jointly utilized with visual features in an early fusion scheme. In \cite{bischke2017detection}, visual features extracted through deep models, pre-trained on ImageNet \cite{deng2009imagenet}, are complemented by textual information, such as users tags, geo-location and temporal information along with textual description. Both textual and visual features are evaluated individually and jointly by concatenating feature vectors. 

Existing pre-trained models are also used in \cite{ahmad2017cnn},  where five different models from four state-of-the-art deep architectures, namely AlexNet \cite{krizhevsky2012imagenet}, GoogleNet \cite{szegedy2016rethinking}, VggNet \cite{simonyan2014very} and ResNet \cite{he2016deep}, pre-trained on the large-scale ImageNet and Places datasets \cite{zhou2014learning}, were used. The basic insight of the paper was to combine object and scene-level features for the flood classification task. Individual Support Vector Machines (SVMs) are then trained on the features extracted through each model, followed by a fusion phase where three different late fusion techniques are used to combine the scores obtained through the individual classifiers along with a Random Forest classifier trained on textual features. 
Object and scene-level features are also used in \cite{ahmad2017convolutional,avgerinakis2017visual,ahmad2018comparative}, for the classification of flooded and non-flooded images in social media. 

Tkachenko et al. \cite{tkachenko2017wisc} rely on hand-crafted visual features, such as the colour and edge directivity descriptor (CEDD) \cite{chatzichristofis2008cedd}, color layout \cite{kasutani2001mpeg} and Gabor wavelets \cite{bai2009novel}. A more sophisticated solution has been proposed for textual information (i.e., description, title and users' tags) relying on word embeddings trained on the entire YFCC100m dataset~\cite{thomee2016yfcc100m}. Each textual feature is extracted separately, and then concatenated to form a single feature vector. Moreover, to translate users' tags into English, a machine translation technique has been employed. In \cite{hanif2017flood}, handcrafted visual features are concatenated into a single feature vector followed by dimensionality reduction and classification phases. Term Frequency Inverse Document Frequency (TFIDF) \cite{salton1988term} are measured for users' tags to represent textual features. In \cite{dao2017domain}, handcrafted visual features along with textual information are used for the classification of flood related images.

An active learning framework intending to collect, filter and analyze social media contents for natural disasters has been proposed in \cite{alam2018processing}. For data collection, a publicly available system, namely AIDR \cite{imran2014aidr}, has been used to crawl social media platforms, followed by a crowd-sourcing activity for data annotation. A pre-trained model \cite{simonyan2014very} is then fine-tuned on the annotated images for classification purposes.

\subsection{Disaster events detection in Satellite Imagery}
Being one of the most valuable sources of information for disaster analysis \cite{ahmad2018social,ahmad2017jord1,klemas2014remote}, a growing portion of research also aims at the detection and classification of natural disaster events in satellite imagery. 
Liu et al.~\cite{liu2016geological} proposed a deep architecture along with a wavelet transformation-based pre-processing scheme for the identification of disaster affected areas in satellite imagery. Amit et al.~\cite{amit2016analysis} propose a CNN-based deep architecture composed of five weighted layers for landslides and flood detection in satellite imagery. 

Benjamin et al. \cite{bischke2017detection} approach the flood detection in satellite imagery as an image segmentation task where a CNN-based framework with three different training strategies has been adopted. In order to remove a location bias due to local changes in images due to lighting conditions and other atmospheric distortions, the individual components of the provided satellite imagery, i.e., RGB and IR, are normalized before training the model. 
In \cite{nogueira2018exploiting}, authors exploit the diversity of different CNNs, which are mainly based on dilated convolution \cite{yu2015multi} and de-convolution \cite{badrinarayanan2017segnet}, in a fusion framework . Initially, binary maps obtained with the individual models are concatenated, which are then used to train SVMs for analyzing which and when the individual model is better. Finally, SVMs are trained on features/maps obtained with the combination of the best models to predict the final binary maps of the test images.

Ahmad et al. \cite{ahmad2017cnn} tackle flood detection in satellite imagery as a generative problem where an Adversarial Generative Networks (GANs) based framework has been proposed. The framework mainly relies on a GANs architecture, namely V-GAN \cite{son2017retinal}, originally developed for the retinal vessel segmentation. In order to adopt the architecture for the flood detection task, the top layer of the generative network is extended with a threshold mechanism to generate binary segmentation mask of the flooded regions in satellite imagery. In an other work from the same authors \cite{ahmad2018social}, the input layer is modified to support 4 channel input images (i.e., RGB and IR) and several experiments are conducted to evaluate the performance of RGB and IR components individually and jointly. 
In \cite{tkachenko2017wisc}, different indices, namely Land Water Index (LWI), Normalised Difference Vegetation Index (NDVI) and Normalised Difference Water Index (NDWI) are selected from the spectral images.  Subsequently, two different strategies based on supervised classification and un-supervised clustering techniques are then adopted for the identification of flooded regions in satellite imagery. On the other hand, Avgrinak et al. \cite{avgerinakis2017visual} rely on Mahalanobis distance~\cite{de2000mahalanobis} and some morphological operations for the task. 

\subsection{MediaEval 2018 challenge on Multimedia Satellite Task: Emergency Response for Flooding Events}

The problem posted in the MediaEval 2018 challenge is different from the existing state-of-the-art, and focuses on a different prospective of detecting road passability in flood affected areas from social media and satellite imagery. In this section, we will briefly discuss the other approaches proposed for the benchmark competition. 

Similar to the challenge posted in MediaEval 2017 \cite{bischke2017multimedia}, in MediaEval 2018, two tasks have been included in the  challenge. The majority of the solutions proposed for the challenge rely on deep architectures. For instance, Fen et al. \cite{Feng2018multimediasatellite} use several deep models pre-trained on Imagenet dataset along with textual features extracted through \textit{fasttext} \cite{bojanowski2017enriching} for the FCSM task. In \cite{lura2018multimediasatellite}, the performance of several deep models is evaluated in a framework with double ended classifier and compact loss function treating the two sub tasks as one class classification problem by individually training the models on the provided dataset. Moreover, data augmentation techniques are used to increase the number of training samples. On the other hand, textual features are represented through embedding initialized with Glove \cite{pennington2014glove}. Another deep architecture based method is presented in \cite{Armin2018multimediasatellite}, where an existing deep model pre-trained on ImageNet is fine-tuned on the provided dataset. Moreover, the Bag of visual Words (BoW) model over the textual information is also used for the FCSM task. For the FDSI task, image patches each of size $50\times 50$ pixels are extracted around each of the two given points. Visual features are then extracted through RGB histograms with 16 bins per channel followed by training and SVMs based classification of the test patches. 

In \cite{Anastasia2018multimediasatellite}, a deep architecture based framework containing two separate deep architectures (VggNet) for the evidence and passability sub-tasks has been proposed in the FCSM task. Each image is passed through the both networks aiming to predict evidence for the road passasbility and differentiating in passable and non-passable images. Moreover, stacked auto-encoders are used for the early fusion of textual and visual features. For the FDSI task, a ResNet model pre-trained on ImageNet is fine-tuned on the satellite imagery provided for the challenge. The authors in \cite{Zhao2018multimediasatellite}, utilize scene-level features extracted through a deep model pre-trained on places dataset for the evidence of road passability. For the second sub-task to differentiate in passable and non-passable roads, both object and scene-level features are extracted through deep models. Moreover, presence of a boat is also used as an indication of the non-passable roads. Hanif et al. \cite{Hanif2018multimediasatellite} adopted an ensemble framework to jointly utilize local features and global visual features for the FCSM task. Global features are extracted through several features descriptors, while a CNN-based local feature descriptor is used for the extraction of local features. On the other hand, textual information is represented through  frequency–inverse document frequency (TF-IDF). Subsequently, an ensemble framework is used to jointly utilize these features for the FCSM task.

\section{Proposed Solutions}
\label{sec:proposed}
In this section, we present our proposed solutions for the both tasks. First, we describe the methodology proposed for the FCSM task, and then we provide the details of the methodology adopted for the FDSI task. 

\subsection{Methodology for FCSM Task}
\begin{figure}[ht!]
\centering
\includegraphics[width=0.56\linewidth]{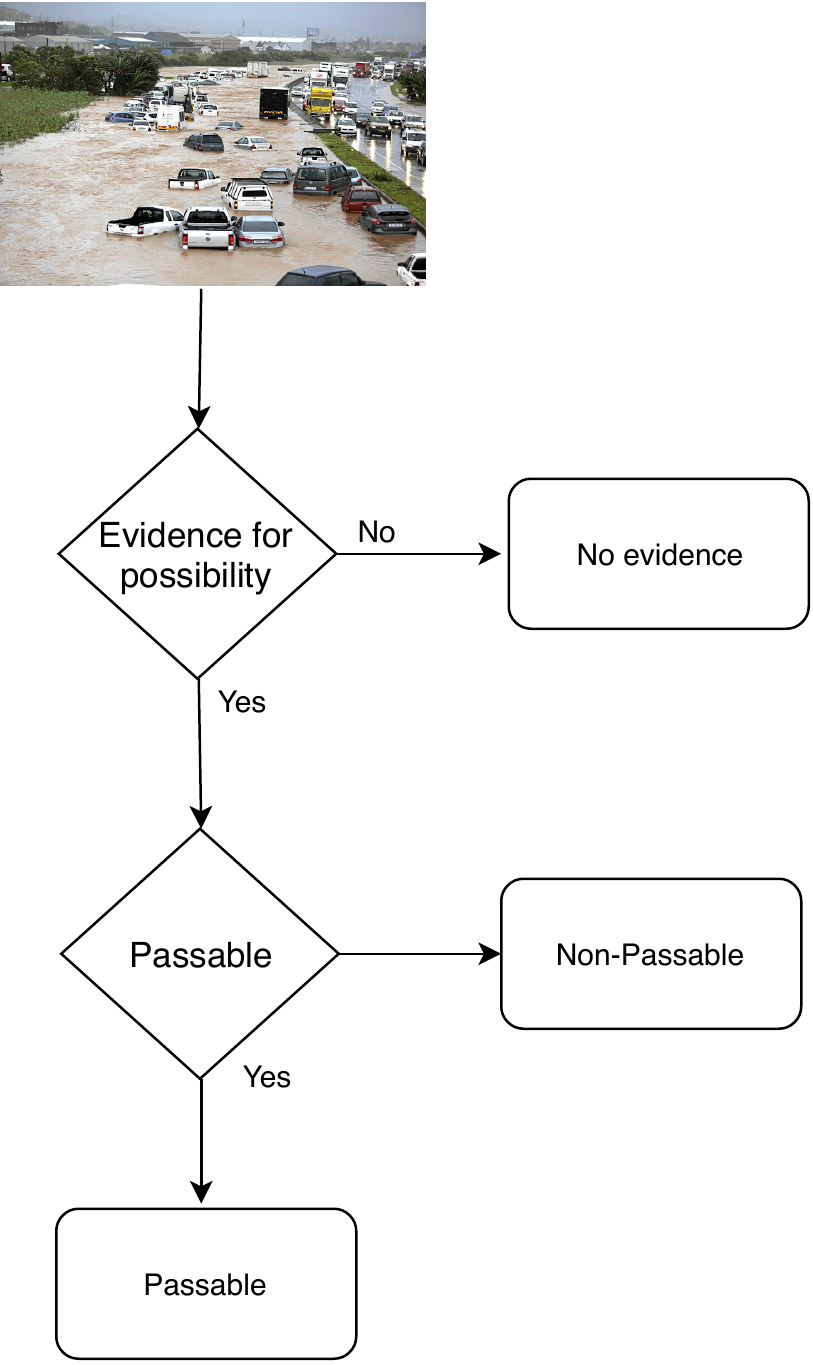}
\caption{Flow chart of the first task showing the two sub-tasks of the FCSM task.}
	\label{flowchart_FCSM}
\end{figure}

The first task is to analyze images from social media providing  direct evidence for passability of roads through conventional means without needing boats and big vehicles, such as trucks. The task can be divided into two sub-tasks, namely (i) identification of images providing an evidence for road passability and (ii) differentiating passable and non-passable images among the ones providing an evidence for the road passability identified in the first step. Both steps are carried out sequentially.
In Figure \ref{flowchart_FCSM}, we provide the flowchart of the first challenge showing the two sub-tasks of the FCSM task. 

\begin{figure*}[ht!]
\centering
\includegraphics[width=.98\linewidth]{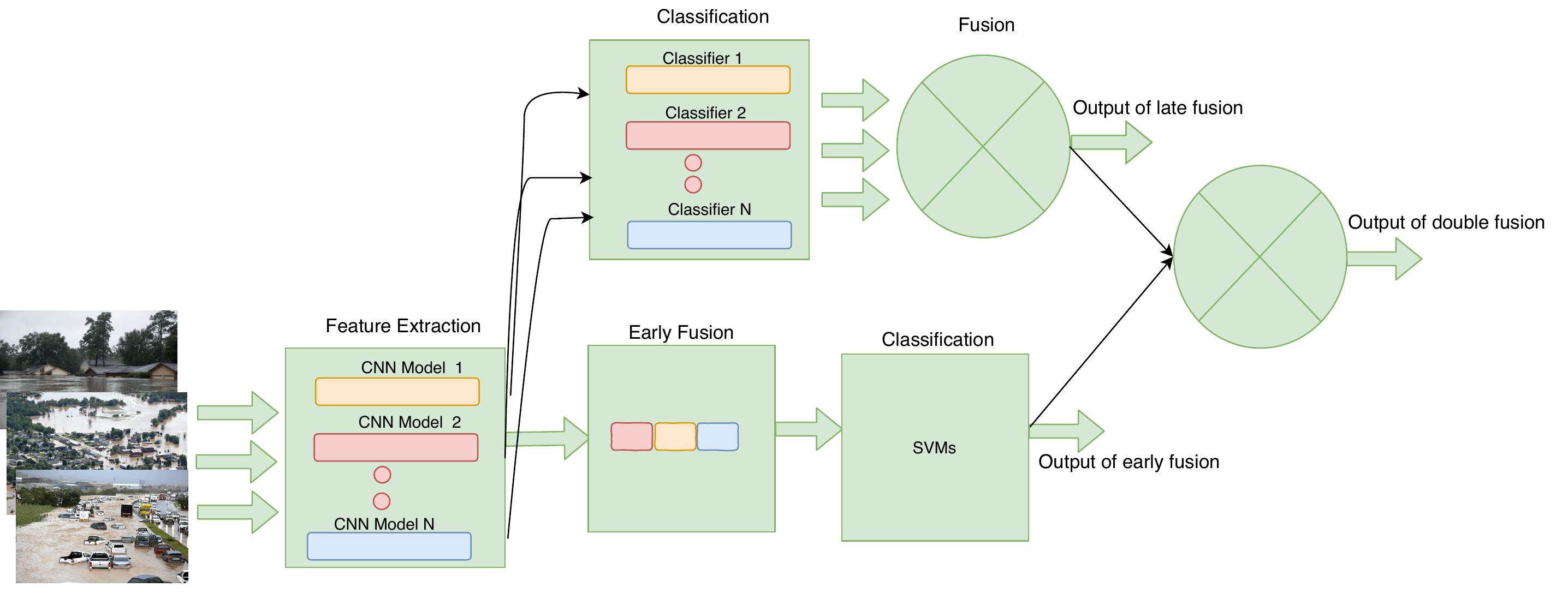}
\caption{Block diagram of the proposed methodology of the three different techniques used for the FCSM task. }
	\label{methodology_FCSM}
\end{figure*}

Figure \ref{methodology_FCSM} shows the block diagram of the methodology we adopted for the FCSM task. The proposed method is mainly composed of three phases, namely, (i) feature extraction, (ii) classification, and (iii) fusion. For feature extraction, we rely on four different models, pre-trained on ImageNet and Places datasets, from three state-of-the-art deep architectures, namely AlexNet \cite{krizhevsky2012imagenet}, VggNet \cite{simonyan2014very} and ResNet \cite{he2016deep}. For classification we adopt a SVM. The basic motivation for the feature extraction through these deep models pre-trained on object and places datasets comes from our previous experience \cite{ahmad2018comparative,ahmad2018ensemble}, where object and scene-level features showed better performance when jointly utilized. In the final phase, we use three different fusion schemes to combine the capabilities of the four models in the FCSM task. In the next subsections, we provide a detailed description of each of the components of the methodology.

\subsubsection{Feature Extraction and classification}
In this phase, we rely on four different deep models for feature extractions. 
Two of the models (ResNet and VggNet) are pre-trained on ImageNet \cite{deng2009imagenet}, and the other two are pre-trained on the Places dataset \cite{zhou2014learning}. 

Features are extracted from the last fully connected layer of each model (i.e., Fc-7 for AlexNet and VggNet, Fc-1000 for ResNet). For VggNet, we use the architecture with 19 layers, and for ResNet the configuration with 50 layers. 
We use the models as feature descriptors without any re-training and fine-tuning. Moreover, for feature extraction with all models, we used the Caffe toolbox\footnote{http://caffe.berkeleyvision.org/}. After feature extraction, SVMs are trained on features extracted through each individual model. 
For classification, we use the default parameters using \textit{Fit-Multiclass} model from the MathWorks\footnote{https://www.mathworks.com} toolbox.

\subsubsection{Fusion}
 To jointly utilize the capabilities of the individual models in the FCSM task, we rely on three different fusion techniques: early, late and double fusion. In the early fusion, we concatenate the features extracted through the different models. For the late fusion, in the current implementation, we simply average the results obtained through the individual models. For the third fusion technique, we combine the results obtained from the first two techniques in an additional late fusion step by averaging their scores. 

\subsection{Methodology for FDSI Task}
For the FDSI sub-task, we have also opted for a CNN and a transfer learning-based classification approach, previously validated in a different application domain in our work~\cite{pogorelov2017holistic}.
In fact, we initially tried to apply the well-performing GAN approach introduced in our previous works for the flood detection satellite imagery~\cite{ahmad2018social} and medical imagery~\cite{pogorelov2018deeppolyp,pogorelov2018deepangie}. 
We conducted an exhaustive set of experiments, but we unfortunately could not achieve a roads passability detection performance better than random label assignment would achieve. The reason for that is the limited size of the dataset (only 1,437 samples were provided in the development set). 
This, in combination with the large variety of landscapes, road types, types of obstacles and weather conditions, etc., prevents the GAN-based approach from adequate training and finding key visual features required to reliably distinguish between flooded and non-flooded roads.

\begin{figure}[ht!]
\centering
\includegraphics[width=0.8\linewidth]{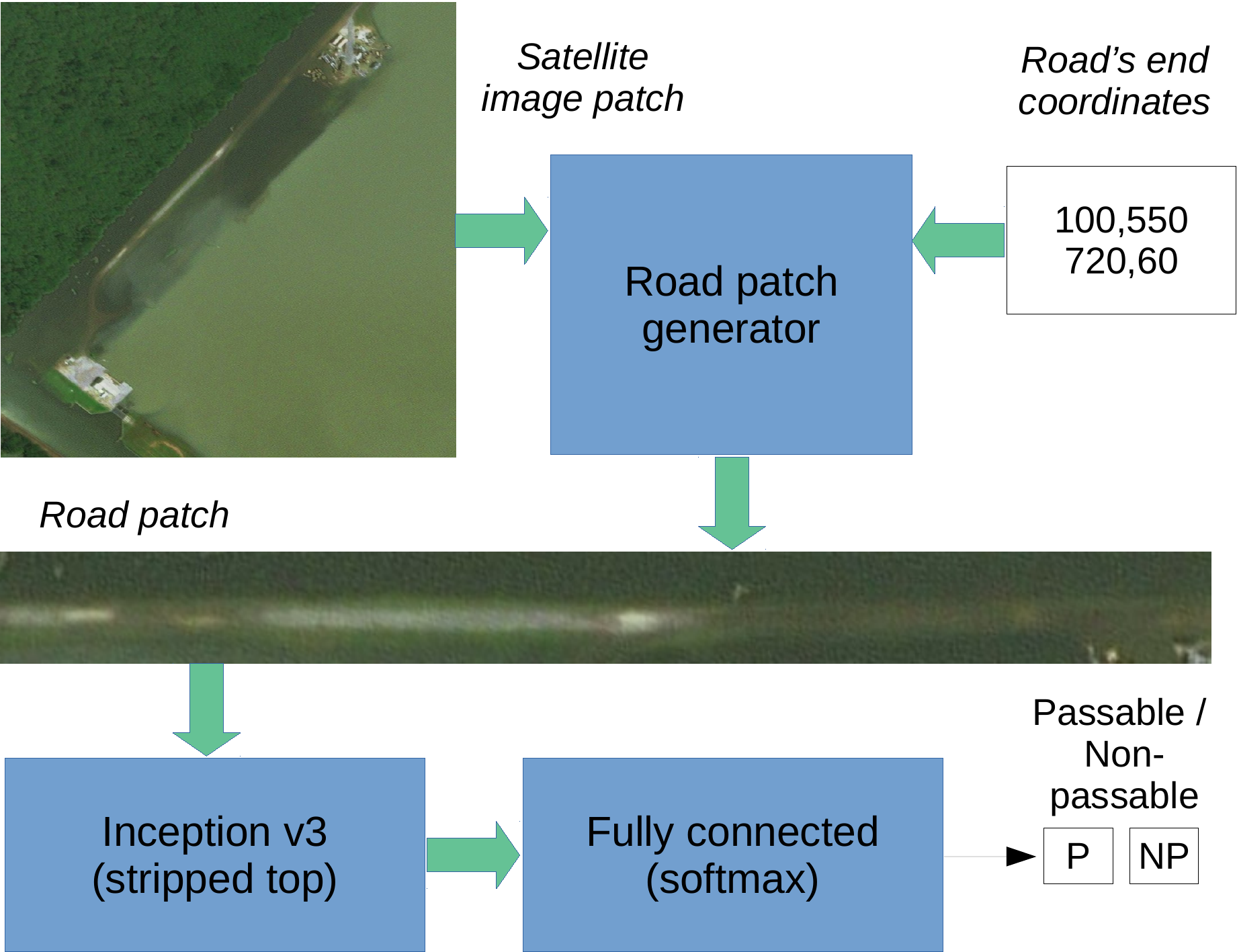}
\caption{Block diagram of the proposed satellite image processing methodology used for the FDSI task.}
	\label{methodology_FDSI}
\end{figure}

Figure \ref{methodology_FDSI} provides the block diagram of the methodology proposed for the FDSI task. This approach is based on the Inception v3 architecture~\cite{szegedy2016rethinking} pre-trained on the ImageNet dataset~\cite{deng2009imagenet} and the retraining method described in~\cite{donahue2014decaf}. 

For the here presented work, we froze all the basic convolutional layers of the network and only retrained the top fully connected layer with softmax activation after random initialization of its weights. The new fully connected layer was retrained using the RMSprop~\cite{tieleman2012lecture} optimizer, which allows an adaptive learning rate during the training process.

As the input for the CNN model, we used the image patches, extracted from the full images using the provided coordinates of the target road end points. Visual inspection of the generated roads' patches from the training dataset, showed relatively good coverage for the road-related areas and enough coverage of the neighbourhood areas and give enough visual information for the following CNN-based analysis and classification.

Moreover, in order to increase the number of training samples, we also performed various augmentation 
operations on the images. Specifically, we performed horizontal and vertical flipping, and change of brightness in the interval of $\pm40\%$.

After the model has been retrained, we used it as a multi-class classifier that provides the probability value for each of two classes: passable and non-passable. The final passability detection is done via the selection of the class with a higher probability. In case of equal class probabilities, we mark the road patch as non-passable.

\section{Evaluation and  Results}
\label{sec:results}
\subsection{Dataset}
The dataset for evaluation is provided in the MediaEval 2018 benchmark competition on Multimedia and Satellite task. In the challenge two different collections containing images and associated meta-data from social media and satellite imagery have been provided for the FSCM and FDSI tasks, respectively. 

For the FCSM task, participants were provided with a collection of tweets along with associated images from three hurricanes, namely Harvey, Maria, and Irma, occurred in 2017. The dataset is downloaded during the events from Twitter using keywords, such as flooding and floods. The dataset also contains additional information, such as rainfall and climate predictions. The development and test sets are provided, separately. The development set is composed of 7,387 tweets and associated images while the test set contains a total of 3683 tweets/images. The ground truth is provided in two separate files, one for each of the sub task (i.e., evidence and passability). Moreover, the participants are also provided with a set of handcrafted visual feature. 

Satellite image patches of flooded areas from the three disaster events have been provided for the FDSI task. The FDSI dataset covers image patches from DigitalGlobe taken by the satellite WorldView3 (0.3m resolution). The dataset is provided in two files, i.e., the one containing the cropped satellite images of flooded areas, while the other provides the ground-truth labels for road passability given two points on each satellite image. The dataset is provided in two sets, namely development and test sets, containing 1,438 and 225 image patches, respectively.
\subsection{Experimental results of the FCSM task}
This section provides a detailed description of the conducted experiments, the results achieved, and their description and comparisons against the state-of-the-art on the FCSM task. 


Table \ref{validation_results} provides the experimental results of our first experiment where we evaluate the performance of the individual deep models on both sub-tasks, namely (i) identification of images providing evidence for road passability, and (ii) differentiation between images showing passable vs. non passable roads, in terms of accuracy per class on both sub-tasks.  
The evaluation is carried out on the development set allocating 60\% and 40\% images for training and testing, respectively. Overall, better results are obtained on the first sub-task, i.e., identification of images providing evidence for road passability. However, the performances for differentiating in passable and non-passable roads are generally lower, demonstrating a significantly higher complexity for the task. Although there is no significant difference in the performance of models in identifying the images providing an evidence of road passability, overall, slightly better results are obtained with VggNet pre-trained on the Places dataset. Moreover, in differentiating in passable and non-passable roads, the models pre-trained on the Places dataset outperform the ones pre-trained on the object dataset showing the importance of scene-level features in the task. 

\begin{table}[!h]
\caption{Evaluation of the individual models in terms of accuracy on the validation set.}
\label{validation_results}
\scalebox{0.79}{
\begin{tabular}{|c|c|c|}
\hline
\multirow{2}{*}{\textbf{Models}} & \multicolumn{2}{c|}{\textbf{Performance (accuracy in \%)}} \\ \cline{2-3} 
 & Evidence sub-task& Passability sub-task\\ \hline
AlexNet (places) & 86.19 & 71.00 \\ \hline
VggNet (places) & 86.79 & 71.85 \\ \hline
VggNet (ImageNet) & 86.79 & 69.85 \\ \hline
ResNet (ImageNet) & 85.05 & 69.28 \\ \hline
\end{tabular}}
\end{table}

In order to analyze the performance of the models on individual classes of the dataset, we also provide the results of the models in terms of per-class accuracy in Table \ref{validation_results_evidence}. In both sub-tasks, the accuracy on the negative samples (i.e., images with no evidence and non-passable images) is generally high. Significant variations in the performances of the individual model can been observed on positive samples (i.e., images providing evidence for passability and images of passable roads). The variations in the performances of the individual models provide basis for our second experiment, where we use three different fusion techniques to combine the capabilities of the individual models for the potential improvement in the performances.   

\begin{table*}[!h]
\caption{Evaluation of the individual models in terms of per class accuracy on the validation set.}
\label{validation_results_evidence}
\scalebox{0.9}{
\begin{tabular}{|c|c|c|c|c|}
\hline
\multirow{2}{*}{\textbf{Models}} & \multicolumn{2}{c|}{\textbf{Evidence sub-task(Accuracy in \%)}} & \multicolumn{2}{c|}{\textbf{Passability sub-task(accuracy in \%)}} \\ \cline{2-5} 
 & Evidence class& No-evidence class& Passable class& Non-Passable class\\ \hline
AlexNet (places) &  78.11 & 90.80 & 57.33  & 81.25\\ \hline
VggNet (places) & 80.00 & 90.71& 57.66 & 82.50 \\ \hline
VggNet (ImageNet)  & 79.88 & 90.77&  55.66 & 80.50 \\ \hline
ResNet (ImageNet) & 77.41 & 89.49& 52.00 & 82.25 \\ \hline
\end{tabular}}
\end{table*}



Table \ref{validation_results_fusion} provides the experimental results of our fusion techniques on both sub-tasks in terms of accuracy on the validation set. 
As can be seen, fusion contributes to significantly improving the performances of the individual models. The double fusion method, which combines the results of both early and late fusion, secures a significant improvement of 6.03\% and 3.71\% over the best single fusion methods, on evidence and passability sub-tasks, respectively.  

\begin{table}[]
\caption{Evaluation results of the fusion experiment in terms of accuracy on the validation set.}
\label{validation_results_fusion}
\scalebox{0.79}{
\begin{tabular}{|c|c|c|}
\hline
\multirow{2}{*}{\textbf{Methods}} & \multicolumn{2}{c|}{\textbf{Performance (accuracy in \%)}} \\ \cline{2-3} 
 & Evidence sub-task& Passability sub-task\\ \hline
Early Fusion & 88.81 & 77.00 \\ \hline
Late Fusion & 90.36 & 76.00 \\ \hline
Double Fusion & 96.43 & 80.71 \\ \hline
\end{tabular}}
\end{table}




We also provide comparisons against state-of-the-art in terms of mean F1 Score, used as an official evaluation metric in the benchmark competition. Since, on one side, we are interested in the evidence class in sub-task 1 and on the other side in the passable and non-passable classes in sub-task 2, the mean F1 score is computed as follows:
\begin{equation}
\scalebox{0.67}{
    Mean\_F1 = (F1 (evidence \; AND \; passable) + F1 (evidence \; AND \; non\_passable))/2}
\end{equation}

Table \ref{tab:comparisons} provides the comparisons of our proposed methods against the existing state-of-the-art. It is important to mention that in the benchmark competition teams could submit up to five runs: visual information only for Run 1, textual information only in Run 2, combination of textual and visual features in Run 3, and two Runs (4 and 5) without any restrictions on the modality of the information. 
In our method, we rely on visual information only. Our first Run is based on late fusion while in the fourth Run, we concatenate  the features extracted through the individual models (early fusion). Our final Run is based on the double fusion. 
The conducted experiments show that the proposed solution outperforms the available state-of-the-art in all four configurations. Our best run with double fusion has significant improvement over most of the methods, and achieves comparable results with method proposed by Anastasia et al. \cite{Anastasia2018multimediasatellite}, which shows the significance of combining multiple deep models for the classification purposes.

\begin{table*}[]
\caption{Comparisons against other methods from the benchmarking competition on the FCSM dataset. In the competition up-to 5 runs were allowed. }
\label{tab:comparisons}
\scalebox{0.90}{
\begin{tabular}{|c|c|c|c|c|c|}
\hline
\multirow{2}{*}{\textbf{Methods}} & \multicolumn{5}{c|}{\textbf{Performance (F1 Measure)}} \\ \cline{2-6} 
 & Visual & Textual & Multi-modal & Run 4 (visual) & Run 5 (visual) \\ \hline
Feng et al. \cite{Armin2018multimediasatellite} & 64.35 & 32.81 & 59.49 & 52.16 & 51.59 \\ \hline
Armin et al. \cite{Armin2018multimediasatellite} & 20.00 & 24.00 & - & 17.00  & 35.00  \\ \hline
Anastasia et al. \cite{Anastasia2018multimediasatellite} & 66.65 & 30.17 & 66.43 & 55.12 & 54.48 \\ \hline
Zhao et al. \cite{Zhao2018multimediasatellite} & 63.88 & 12.86 & - & 63.13 & 63.89 \\ \hline
Hanif et al. \cite{Hanif2018multimediasatellite} & 45.04 & 31.15 & 45.56 & - & - \\ \hline
Our method & 63.58 (late fusion) & - & - & 60.59 (early fusion) & 65.03 (double fusion ) \\ \hline
\end{tabular}}
\end{table*}

\subsection{Experimental results of the FDSI Task}
For the experimental setup of the FDSI task, we decided to perform only two mandatory runs, which rely on the task-provided training data only. Due to a limited amount of training samples available, training the deep architecture from the scratch is not possible. Thus, we decided to perform two types of training for our transfer-learning detection approach. 

First, we implemented a pipeline for classification that differs from common procedures. This process was involving all the training samples into the training process as both training and validation sets. Usually, for classification tasks, this would result into over-fitting of the model and inability to correctly classify the test samples. However, for this specific task, the limited number of training epochs and significant training data augmentation in conjunction with a high variety of road patch samples resulted in normal training process. This allowed to correctly retrain the last layers of the network and produce reasonable classifiers even on such a limited training set. 

The official F1-Score metric (see table~\ref{roads_results}) on the non-passable road class for the first "All-train" Run is $62.30\%$.
To verify our idea of the usability of using all the training data for both training and validation, we also performed a normal network training with a random 50/50 development/validation data split. This second \emph{Half-trained} Run resulted in F1-Score of $61.02\%$ which is slightly lower comparing to the \emph{All-trained} Run. This is confirming the validity of our idea of using the complete training dataset and heavy data augmentation to improve road patches classification performance.

\begin{table}[]
\caption{Evaluation of our proposed approach for the FDSI task in terms of F1 Scores.}
\label{roads_results}
\begin{tabular}{|c|c|c|}
\hline
\textbf{Run} & \textbf{Method} & \textbf{F1 Score} \\ \hline
1 & All-train & 62.30\% \\ \hline
2 & Half-train & 61.02\% \\ \hline
\end{tabular}
\end{table}

We also provide comparisons of our method for FDSI task against state-of-the-art in terms of F1 score in Table \ref{tab:sat_comparisons}. 
Our method outperforms the state-of-the-art in both Run 1 and Run 2 with an improvement of 5.30\% and 4.02\%, respectively. 

\begin{table}[]
\caption{Comparisons against other state-of-the-art methods from the benchmarking competition on FDSI dataset. In the competition, up-to 5 runs were allowed.}
\label{tab:sat_comparisons}
\scalebox{0.80}{
\begin{tabular}{|c|c|c|c|c|c|}
\hline
\multirow{2}{*}{\textbf{Methods}} & \multicolumn{5}{c|}{\textbf{Performance (F1 Measure)}} \\ \cline{2-6} 
 & Run 1 & Run 2 & Run 3 & Run 4  & Run 5  \\ \hline
Armin et al. \cite{Armin2018multimediasatellite} & 57.00 & 32.00 & 39.00 &  56.00 & 57.00  \\ \hline
Anastasia et al. \cite{Anastasia2018multimediasatellite} & 56.45 & - & - & - & - \\ \hline
Our method & 62.30  & 61.02 & - & - & - \\ \hline
\end{tabular}}
\end{table}

\section{Conclusions and Future Work}
\label{sec:conclusion}
In this paper, we addressed a challenging problem of detecting the passibility of roads.
In the social media image analysis, we mainly relied on deep features extracted through different pre-trained deep models individually and jointly through fusion. We observed better results can be achieved for the models pre-trained on Places dataset compared to the ones pre-trained on objects dataset, showing the importance of the scene-level information in the task. We also observed that the object-level information well complement the scene-level features when jointly utilized. 
Among the fusion methods, double fusion combines the capabilities of both early and late fusions and ultimately leads to better results. Considering the improvement with fusion techniques, in future, we aim to use some optimization methods to assign more specific weights to the individual models. 

In the satellite sub-task, we found that just a normal image segmentation approach is of no help, and we implemented a task-oriented CNN and transfer learning-based approach. This approach was able to classify image patches with roads and achieved an F1-Score of $62.30\%$ for the non-passable road class. In the future, we plan to implement an advanced road network and flooding detection and segmentation using a combined CNN- and GAN-based approach pre-trained on the existing annotated road network and flooded areas datasets.

\section*{Acknowledgments}

This research is partly supported by the
ADAPT Centre for Digital Content Technology, which is funded under the Science Foundation Ireland Research Centres Programme (Grant 13/RC/2106) and is co-funded under the European Regional Development Fund.
{\small
\bibliographystyle{ieee}
\bibliography{egbib}
}

\end{document}